\documentclass{elsart}
\begin{document}

\begin{frontmatter}
\title{Comment on ``Evaluation of the Pion-Nucleon Sigma Term from CHAOS Data''}
\author[EM]{E. Matsinos{$^*$}},
\author[GR]{G. Rasche},
\address[EM]{Institute of Mechatronic Systems, Zurich University of Applied Sciences, Technikumstrasse 5, P.O. Box, CH-8401 Winterthur, Switzerland}
\address[GR]{Institut f\"{u}r Theoretische Physik der Universit\"at, Winterthurerstrasse 190, CH-8057 Z\"{u}rich, Switzerland}

\begin{abstract}
The value of $59 \pm 12$ MeV for the pion-nucleon ($\pi N$) $\Sigma$ term, which Stahov, Clement, and Wagner recently extracted from the differential cross sections (DCSs) of the CHAOS Collaboration, does not match well the expectation of an enhanced (more positive) 
isoscalar component in the $\pi N$ interaction at low energies, which the rest of the modern (meson-factory) data favour. However, we have already demonstrated that the angular distribution of the CHAOS $\pi^+ p$ DCSs is not compatible in shape with the rest of the 
modern low-energy $\pi^+ p$ data. This problem must be addressed and resolved by the CHAOS Collaboration prior to the extrapolation of their partial-wave amplitudes into the unphysical region.
\end{abstract}
{$^*$}{Corresponding author. E-mail: evangelos.matsinos@zhaw.ch, evangelos.matsinos@sunrise.ch; Tel.: +41 58 9347882; Fax: +41 58 9357306}
\end{frontmatter}

Stahov, Clement, and Wagner \cite{scw} recently evaluated the pion-nucleon ($\pi N$) $\Sigma$ term from the $\pi^\pm p$ differential cross sections (DCSs) of the CHAOS Collaboration \cite{chaos,denz}; the extracted value was $59 \pm 12$ MeV. In Ref.~\cite{mr1}, which 
has been available online since the beginning of April 2013, we reported the details of a partial-wave analysis (PWA) of the same data. To avoid a bias from extraneous sources, we performed therein an exclusive analysis of the CHAOS DCSs, applying to these data the 
same analysis criteria which were earlier applied to the rest of the low-energy $\pi N$ measurements \cite{mr2}.
\begin{itemize}
\item In the first part of the analysis, we used standard low-energy parameterisations of the $s$- and $p$-wave $K$-matrix elements, thus avoiding to impose the theoretical constraint of crossing symmetry onto the fitted scattering amplitudes. The results of the 
optimisation suggested the removal of a few obvious outliers (eleven degrees of freedom, in total) from the initial CHAOS database of $546$ data points. However, the final results of the optimisation disagreed with the $\pi^- p$ scattering lengths obtained 
(experimentally) from pionic hydrogen at threshold. (Further analysis revealed that this result was due to the inadequacy of the isospin-$\frac{3}{2}$ amplitude to simultaneously account for the $\pi^+ p$ and $\pi^- p$ DCSs of the CHAOS Collaboration.)
\item After the removal of the eleven outliers, we attempted to fit the ETH model~\footnote{The ETH model of the $\pi N$ interaction contains $t$-channel $\sigma$- and $\rho$-exchange graphs, as well as the $s$- and $u$-channel contributions with all the well-established 
$s$ and $p$ baryon states with masses below $2$ GeV; the model obeys crossing symmetry and isospin invariance.} \cite{glmbg} to the combined elastic-scattering database of the CHAOS Collaboration ($535$ degrees of freedom). The ability of this model to account for the 
hadronic part of the $\pi N$ interaction, even above the energy of the $\Delta(1232)$ resonance, has amply been demonstrated during the past two decades. We found that the fitted values of the model parameters were far from those established during the long-term 
application of this model onto the modern (meson-factory) data; furthermore, the evaluation of the correlation (Hessian) matrix of the fit failed (which has the consequence that the output uncertainties of the model parameters, and of all the predictions obtained on their 
basis, cannot be estimated). The $s$- and $p$-wave phase shifts (`central' values, no uncertainties), extracted from the CHAOS data, were found to be incompatible with the results of Refs.~\cite{mr2,mworg,abws}, and their energy dependence was puzzling.
\item To trace the origin of these problems, we subsequently investigated the reproduction of the CHAOS DCSs using the results of our recent PWA \cite{mr2}. We found that the absolute normalisation of the CHAOS $\pi^- p$ data was in good agreement with the corresponding 
predictions of Ref.~\cite{mr2}, as was the normalisation of their $\pi^+ p$ data sets at backward scattering angles. Large effects in the normalisation of the CHAOS $\pi^+ p$ data sets were observed at forward and medium scattering angles. We therefore concluded that the 
angular distribution of the CHAOS $\pi^+ p$ DCSs, at all five energies covered by the experiment, was not compatible in shape with the rest of the modern data (see Fig.~1 of Ref.~\cite{mr1}).
\end{itemize}

There is little doubt that the rest of the modern data support an enhanced (compared to Koch's amplitudes \cite{kk}) isoscalar component in the hadronic part of the $\pi N$ interaction at low energies. As the $\pi N$ $\Sigma$ term is an isoscalar quantity, its estimate 
from the modern data is also expected to exceed Koch's canonical value (of about $60$ MeV). Despite the differences we have with the SAID solution \cite{abws} at low energies, their result for the $\Sigma$ term ($79 \pm 7$ MeV) \cite{pasw} is not unreasonable.

In view of the problems reported in Ref.~\cite{mr1}, we cannot recommend the inclusion of the DCSs of Refs.~\cite{chaos,denz} in sensitive low-energy analyses. The problems surrounding the CHAOS results in the physical region must be addressed and resolved prior to the 
extrapolation of their partial-wave amplitudes into the unphysical one.

\end{document}